\documentclass[twocolumn,showpacs,preprintnumbers,amsmath,amssymb,prb,aps]{revtex4-1}
\usepackage{epsfig}
\usepackage{graphicx}
\usepackage{dcolumn}
\usepackage{bm}
\usepackage{textcomp}
\usepackage{subfig}

\begin{document}

\title{Understanding the High Temperature Thermoelectric Properties of La$_{0.82}$Ba$_{0.18}$CoO$_{3}$ Compound using DFT+U Method}
\author{Saurabh Singh${{^{1}}}$}
\altaffiliation{Electronic mail: saurabhsingh950@gmail.com}
\author{Devendra Kumar${{^{2}}}$}
\author{Sudhir K. Pandey${{^{1}}}$}
\affiliation{${{^{1}}}$School of Engineering, Indian Institute of Technology Mandi, Kamand - 175005, India}

\affiliation{${{^{2}}}$UGC-DAE Consortium for Scientific Research, University Campus, Khandwa Road, Indore-452001,India}

\date{\today}
\begin{abstract}
Normally, understanding the temperature dependent transport properties of strongly correlated electron systems remains challenging task due to complex electronic structure and its variations (around E$_{F}$) with temperature. Here, we report the applicability of DFT+U in explaining thermopower ($\alpha$) and electrical conductivity ($\sigma$) in high temperature region. We have measured temperature dependent $\alpha$ and $\sigma$ in the 300-600 K range. The non-monotonic temperature dependent behavior of $\alpha$ and metallic behavior of $\sigma$ were observed. The value of $\alpha$ at 300 K was $\sim$15.80 $\mu$V/K and it decreases upto $\sim$477 K ($\sim$11.6 $\mu$V/K) and it further increases with temperature to the $\sim$14.8 $\mu$V/K at 600 K, whereas the values of $\sigma$ were found to be $\sim$1.42 $\times$10$^{5}$ $\Omega$$^{-1}$ m$^{-1}$ and $\sim$0.20 $\times$10$^{5}$ $\Omega$$^{-1}$ m$^{-1}$ at 300 and 600 K, respectively.  Combining the WIEN2k and BoltzTraP code, the electronic structure and temperature dependent transport coefficients were calculated. The ferromagnetic ground state electronic structure with half-metallic character obtained from the DFT+U calculations, U = 3.1 eV, provides better explanation of high-temperature transport behavior. Two current model was used for calculation of $\alpha$ and $\sigma$ where the temperature dependent values of relaxation time ($\tau$), almost linear for up-spin, $\tau$$_{up}$, and non-linear for dn-spin, $\tau$$_{dn}$, were used and estimated values were found to be in good agreement with experimentally reported values.

\end{abstract}


\maketitle
\section{Introduction} 
In the past several decades, study of strongly correlated electron systems (SCES) have been very attractive as they show unusual (often technologically useful) properties, like thermopower ($\alpha$, also known as Seebeck coefficient), electrical conductivity ($\sigma$), colossal magnetoresistance, metal-insulator transition, high temperature superconductivity, half-metallicity, and etc.\cite{Tokura, Imada, Antonov, Pruschke} Most of these properties have been found in the transition metal compounds due to strong interplay between charge, spin, and orbital degrees of freedom. Among these, spin and orbital degrees of freedom have crucial role in tuning the $\alpha$ values.\cite{Koshibae, Koshiba, Maekawa} The external parameter such as temperature also affect the $\alpha$ behavior, and it often shows the non-monotonic temperature dependence.\cite{Oudovenko, Uchida, Matsuo}\\
The different fascinating physical properties possessed by SCES systems have been explored through numerous experimental and theoretical tools. The Density functional theory (DFT) based tools have been contributed significantly in understanding of material properties at the microscopic level.\cite{Sholl, SinghDJ} However, it fails to describe the electronic structure and physical properties of the electron systems in which the interaction among the electrons is strong, particularly where physical properties arises from the correlations among \textit{3d} and \textit{4f} electrons. For such systems, advanced methods of electronic structure determination such as LSDA plus self-interaction corrections (SIC-LSDA), the DFT+U method, the GW approximation, and dynamical mean-field theory have met the considerable success.\cite{JPPerdew, Ansimov, Hedin, Georges} Among these methods, DFT+U method is simplest, cost effective, and being used most frequently from last two decades. The DFT+U tools are based on the static model and it generally not considered in the study of those systems where temperature dependent evolution of electronic structure take place. For some SCES having metal, half-metal, and semi-metal ground state electronic structure, the temperature dependent change in density of states (DOS) have been seen.\cite{Bindu, Lombardo, Ebata} In such kind of systems, with change in temperature a significant changes in DOS around Fermi level, E$_{F}$ (in \textit{k$_{B}$}T range), is more effective in deciding the values of $\alpha$ and $\sigma$, as charge carriers in \textit{k$_{B}$}T range around E$_{F}$ are main contributors in the values of these transport coefficients. In this context, DFT+U tools are expected to be fail in describing the high temperature behavior of $\alpha$ and $\sigma$. However, in our earlier study DFT+U have been found successful in explaining the $\alpha$ behavior of LaCoO$_{3}$, ZnV$_{2}$O$_{4}$ and La$_{0.75}$Ba$_{0.25}$CoO$_{3}$ compounds in 300-600 K range. In case of LaCoO$_{3}$ and ZnV$_{2}$O$_{4}$, consideration of appropriate values of temperature dependent gap made a phenomenal role in the analysis of $\alpha$ data, but this approach is limited for the systems having insulating ground state. For La$_{0.75}$Ba$_{0.25}$CoO$_{3}$ system, the suitable values of temperature dependent $\tau$ were taken into account for estimation of $\alpha$ and have been found in good agreement with experimental data. These studies were limited for explaining the only one physical parameters i.e. $\alpha$. In order to see the versatility of this method in the analysis of high temperature TE behavior, we have chosen another system i.e. La$_{0.82}$Ba$_{0.18}$CoO$_{3}$. It will be also interesting to see whether DFT+U approach and suitable values of temperature dependent $\tau$ can be capable in explaining the $\alpha$ and $\sigma$ behavior of this system in high temperature region. To the best of our knowledge, high temperature TE behavior of La$_{0.82}$Ba$_{0.18}$CoO$_{3}$ have not been explored so far.\\
In the present work, we have measured the temperature dependent $\alpha$ and $\sigma$ in the temperature range 300-600 K. The electronic structure calculation shows half-metallic FM ground state of the system with an energy gap $\sim$270 meV for dn-spin channel. The suitable values of temperature dependent $\tau$ for up and dn-channel are used in the two current model and estimated values of $\alpha$ are found to be closer to the experimental data. A good match between experimental and calculated values of $\alpha$ is found. The estimated values of $\sigma$ by using the same values of $\tau$, which used to calculate the values of $\alpha$, are found to be similar to the observed data. Thus, in the present study  we found that the DFT+U tools not only explain the $\alpha$ data but also give better explanation of the $\sigma$ behavior in 300-600 K range. 
\section{Experimental and Computational details }
We have prepared polycrystalline La$_{0.82}$Ba$_{0.18}$CoO$_{3}$ cobaltite through pyrophoric method.\cite{Pati} Synthesis of the present sample under study is similar to the earlier synthesized compounds, where different amount of barium doping at lanthanum site of LaCoO$_{3}$ have been done. The detailed synthesis information are provided in the previous work carried out by Devendra \textit{et al}.\cite{Devendrab, Devendra} The as-prepared powders were calcined at 1125 $^{0}$C for 12 h. To perform the $\alpha$ and $\sigma$ measurements, the powders were further pelletized under the pressure of $\sim$35 kg/cm$^{2}$ and sintered at 1125  $^{0}$C for 12 h. $\alpha$ measurement were carried out on the pellet having the diameter and thickness $\sim$5 mm and $\sim$0.5 mm, respectively. The resistivity measurement were performed on the rectangular bar shaped sample with dimensions of $\sim$5 mm length, $\sim$1 mm width and $\sim$0.5 mm thickness by using four-probe method. The inverse of resistivity were taken to obtain the values of $\sigma$. Both the transport characterizations were carried out in 300-600 K temperature range using the home-made set up.\cite{Singh}\\
The electronic and TE properties of the compound have also been studied by combine use of the full potential linearized augmented plane-wave (FP-LAPW) method implemented in WIEN2k code and BoltzTraP code.\cite{Blaha, Madsen} Temperature dependent aspects in calculations of transport coefficients ($\alpha$ \& $\sigma$) are taken through the Fermi Dirac distribution function and the detailed of this can be found in Ref. [28].  The exchange correlation function within local density approximation (LSDA) of Perdew and Wang is used.\cite{Perdew} The self consistent field calculations corresponding to non-magnetic and magnetic phase were carried out by using the virtual crystal approximation method. A conventional unit cell having two formula unit were taken, from that an amount of 0.36 electrons were removed and the remained unit cell was assumed as an equivalent to the La$_{0.82}$Ba$_{0.18}$CoO$_{3}$. The Muffin-Tin sphere radii for La, Co and O atoms were fixed to 2.46, 1.97 and 1.69 Bohr, respectively. The value of on-site Coulomb
interaction strength, U, was taken equal to 3.1 eV and found to be explaining the transport data in better way. The convergence criteria in the self-consistency iteration were set to be 10$^{-6}$ Ry and 10$^{-3}$ electronic charge for the total energy/cell and charge/cell, respectively. The lattice parameters (a = 5.4549 $\AA$ and c = 13.3194 $\AA$) of rhombohedral structure described by space group R$\bar3$c were used for the calculation. The value of \textit{R$_{MT}$K$_{max}$} parameter was set to 7, whereas for electronic properties and transport coefficient calculations the values of \textit{k}-integration mesh and \textit{lpfac} parameter were chosen to 50 $\times$ 50 $\times$ 50 and 5, respectively.
\section{Results and Discussion}
Fig. 1 shows the temperature dependent $\alpha$ data in 300-600 K temperature range. The positive values of $\alpha$ are observed in the entire temperature range under study. 
\begin{figure}[htbp]
  \begin{center}
   \vspace{1.2cm}
   \includegraphics[width=0.35\textwidth]{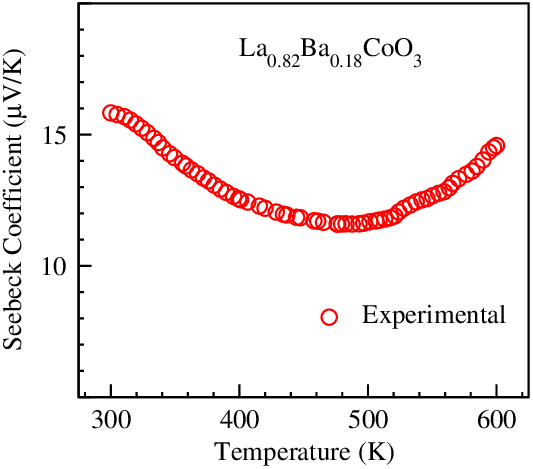}
    \label{}
    \captionsetup{justification=raggedright,
singlelinecheck=false
}
    \caption{(Color online) Temperature dependent Seebeck coefficient $\alpha$ for La$_{0.82}$Ba$_{0.18}$CoO$_{3}$ compound.}
    \vspace{-0.3cm}
  \end{center}
\end{figure} 
The value of $\alpha$ at 300 K is $\sim$15.8 $\mu$V/K and found to be almost equal to the value reported for La$_{0.80}$Ba$_{0.20}$CoO$_{3}$ by Mandal et al.\cite{Mandal} The values of $\alpha$ decreases very slowly up to $\sim$477 K ($\sim$11.6 $\mu$V/K). Above 477 K, the increment in the values of $\alpha$ are noticed and it reaches to $\sim$14.8 $\mu$V/K at 600 K.  The continuous decrease in the values of $\alpha$ in the 300-477 K range is a signature of metallic nature of the system, whereas further increase in the values of $\alpha$ above 477 K can be possible due to the competing contributions from up and dn-channel of half-metallic system and results into the net increment in the contributions of $\alpha$ of the system. In our earlier study, half-metallic nature is found in similar system i.e. La$_{0.75}$Ba$_{0.25}$CoO$_{3}$.\cite{Saurabh}  For present system (La$_{0.82}$Ba$_{0.18}$CoO$_{3}$) in the present study, half-metallic nature with more energy band gap in the vicinity of Fermi level (for dn-spin channel) are expected. In comparison to La$_{0.75}$Ba$_{0.25}$CoO$_{3}$, larger values of $\alpha$ are noticed in the La$_{0.82}$Ba$_{0.18}$CoO$_{3}$ in the entire temperature range.\\
Fig. 2 shows the temperature dependent $\sigma$ data in 300-600 K range. The value of $\sigma$ at 300 K is $\sim$1.42 $\times$10$^{5}$ $\Omega$$^{-1}$ m$^{-1}$. The non-linear decrease in the values of $\sigma$ with T is noticed in the entire temperature region under study. The value of $\sigma$ at $\sim$600 K is found to be $\sim$0.20 $\times$10$^{5}$ $\Omega$$^{-1}$ m$^{-1}$, which is almost 7 times smaller than that of found at 300 K. The temperature dependent variations of $\sigma$ shows metallic nature. To understand these transport behavior of the compound, we have further performed the electronic structure as well as temperature dependent transport coefficients calculations.
\begin{figure}[htbp]
  \begin{center}
  \vspace{0.9cm}
   \includegraphics[width=0.35\textwidth]{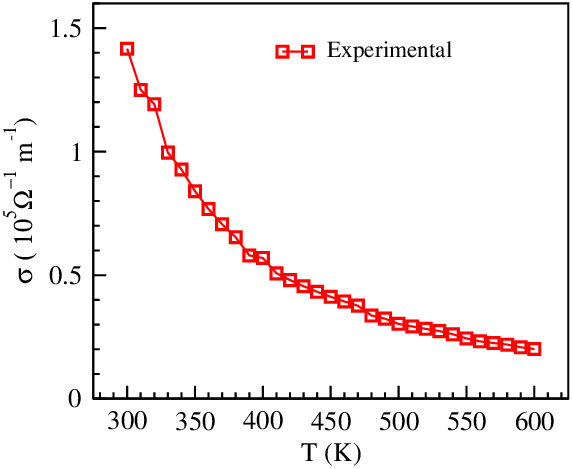}
    \label{}
    \captionsetup{justification=raggedright,
singlelinecheck=false
}
    \caption{(Color online) Temperature dependence of $\sigma$ for La$_{0.82}$Ba$_{0.18}$CoO$_{3}$ compound.}
    \vspace{-0.3cm}
  \end{center}
\end{figure} 

In order to know the magnetic ground state of this system, we have carried out the self consistency calculations for both non-magnetic and ferromagnetic (FM) phase. For the FM solution, the value of total converged energy is found to be $\sim$122 meV/f.u. lower than that of the value for non-magnetic solution. This clearly suggest that ground state of this system is FM, which is also in accordance with experimental report.\cite{Kriener} Therefore, the electronic structure calculations were performed corresponding to the FM phase.

\begin{figure}[htbp]
  \begin{center}
   \includegraphics[width=0.35\textwidth, totalheight=0.45\textheight]{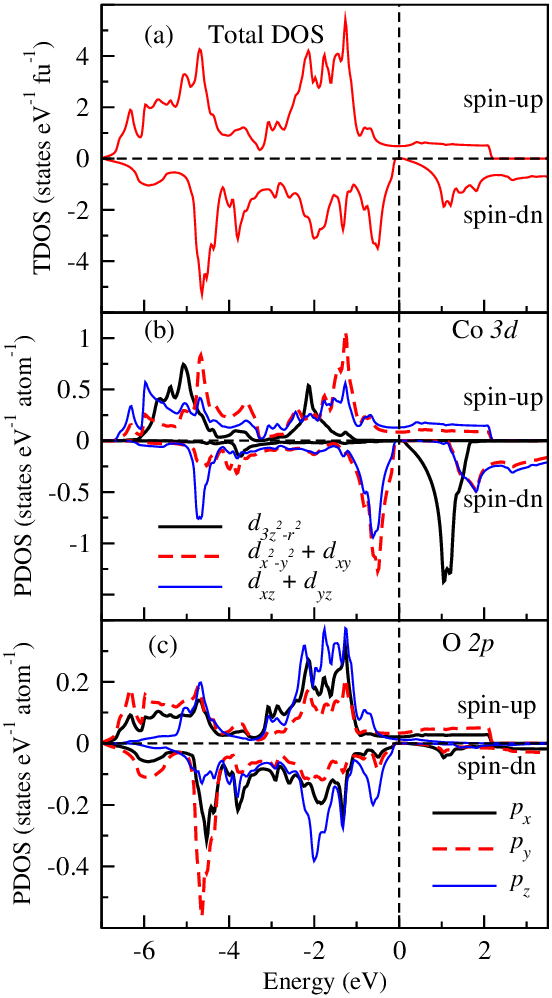}
    \label{}
    \captionsetup{justification=raggedright,
singlelinecheck=false
}
    \caption{(Color online) Total and partial density of states plots for La$_{0.82}$Ba$_{0.18}$CoO$_{3}$. Shown are (a) the TDOS plot, (b) PDOS of Co atom (\textit{3d} orbitals), (c) PDOS of O atom (\textit{2p} orbitals).}
    \vspace{-0.5cm}
  \end{center}
\end{figure}
The Total density of states (TDOS) plot for FM phase is shown in Fig. 3a. The dashed line at 0 eV represents the Fermi level (E$_{F}$). The value of DOS at E$_{F}$ is $\sim$0.48 states/eV/f.u. for the spin-up channel, whereas for the dn-spin channel an energy band gap of $\sim$270 meV is found in the vicinity of Fermi level. It clearly shows that this system is half metallic. From the TDOS plot of this system it has also been noticed that the value of DOS at E$_{F}$ for up spin is $\sim$0.02 states/eV/f.u. smaller and energy band gap for dn spin is $\sim$220 meV larger than that have been noticed in the case of La$_{0.75}$Ba$_{0.25}$CoO$_{3}$. In comparison to La$_{0.75}$Ba$_{0.25}$CoO$_{3}$, the DOS feature at E$_{F}$ of this system clearly suggests that the contributions in $\sigma$ from up and down channel will be smaller, where as the large contributions in $\alpha$ will be from the dn-channel. The energy band gap for dn-channel is almost half of the value obtained for the insulating parent, LaCoO$_{3}$ ($\sim$0.5 eV), compound.\cite{Singh} Therefore, in this system a large contributions in the values of $\alpha$ will be from the dn-channel. To see the contributions in transport properties from different atomic states around E$_{F}$, we have also calculated the partial density of states (PDOS) for Co and O atom. PDOS plots for Co \textit{3d} and O \textit{2p} orbitals are shown in Fig. 3b and 3c, respectively. It is evident from the Fig. 3b that for up-spin the main contributions in DOS at E$_{F}$ is from \textit{d$_{xz}$} + \textit{d$_{yz}$} and \textit{ d$_{x^{2}-y^{2}}$} + \textit{d$_{xy}$} orbitals, and its values are $\sim$0.128 and $\sim$0.08 states/eV/atom, respectively. For dn-spin, at the edge of valence band a DOS peak associated with each of \textit{ d$_{x^{2}-y^{2}}$} + \textit{d$_{xy}$} and \textit{d$_{xz}$} + \textit{d$_{yz}$} orbital is found at $\sim$-0.5 and $\sim$-0.61 eV, respectively. The values of DOS are $\sim$-1.27 and $\sim$-0.94 states/eV/atom, respectively. This shows that the fraction of thermally excited electrons from VB to CB will be maximum from \textit{ d$_{x^{2}-y^{2}}$} + \textit{d$_{xy}$} and \textit{d$_{xz}$} + \textit{d$_{yz}$} orbitals of dn-spin of the Co \textit{3d} atoms. These thermally excited electrons will be mainly contribute in the transport properties of the system. The PDOS plots for \textit{p$_{x}$}, \textit{p$_{y}$} and \textit{p$_{z}$} of O \textit{2p} orbitals in Fig. 3c clearly shows that there are negligible small DOS from \textit{p$_{x}$}, \textit{p$_{y}$} and \textit{p$_{z}$} at E$_{F}$, where as small DOS for \textit{p$_{z}$} equal to $\sim$-0.20 states/eV/atom is found at $\sim$-0.61 eV in VB. This suggests that for temperature range under study, there will be negligible small contributions in transport properties from the O \textit{2p} orbitals. In the system showing half-metallic character, the consideration of both the spin channel contributions in the transport properties is necessary and it is also essential to estimate the total values of transport coefficients from up and down-spin channels. For the half-metallic system, the dn-spin channel have a small energy gap around E$_{F}$ and shows semiconducting nature where both types of charge (holes and electrons) carriers have different contributions in transport behavior due to their different effective masses and mobility. Therefore, the calculations of dispersion curve plays an important role in the qualitative understanding of the transport properties.\\
The dispersion curves plots along the high symmetric points ($\Gamma$-T-L-$\Gamma$-FB-T) for spin-up and spin-dn channels are shown in Fig 4a and 4b, respectively. Two bands (\textbf{1} $\&$ \textbf{2}) are highly dispersive and crosses the Fermi level 10 times. The electrons from these two bands have main contribution in the $\sigma$. The nature of bands crossover the E$_{F}$ are almost similar to those found in case of La$_{0.75}$Ba$_{0.25}$CoO$_{3}$. However, a small shift of $\sim$150 meV in the band lines towards the VB is noticed. Thus, the contributions from up-channel in the values of $\sigma$ and $\alpha$ are expected to small in comparison to La$_{0.75}$Ba$_{0.25}$CoO$_{3}$. For dn-channel an indirect band gap of $\sim$270 meV is also found.

  \begin{figure}[htbp]
    \captionsetup[subfigure]{labelformat=empty}
\begin{center}

\subfloat[]{
        
        \includegraphics[clip,trim=0.cm 0cm 0.0cm 0.0cm, width=0.35\textwidth]{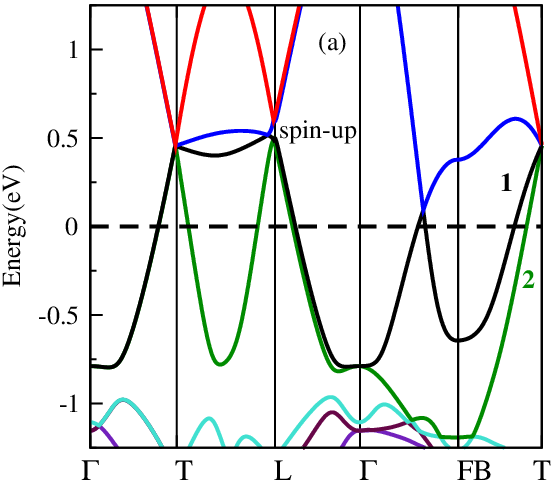} } 
\vspace{-0.40cm}
\subfloat[]{
       
        \includegraphics[clip,trim=0cm 0.0cm 0cm 0.0cm, width=0.35\textwidth]{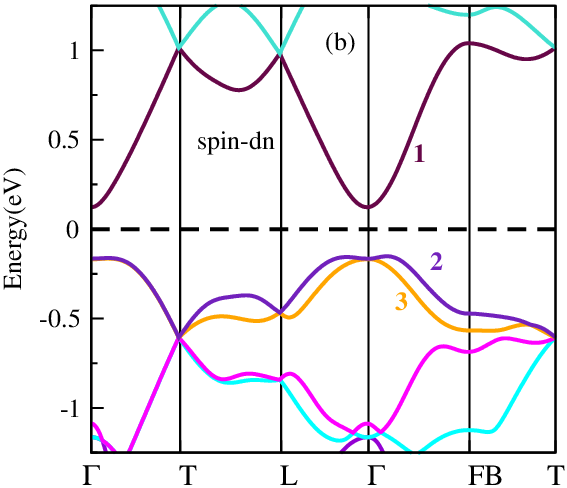} } 
\captionsetup{justification=raggedright,
singlelinecheck=false
}
\caption{(Color online) Electronic band structure of La$_{0.82}$Ba$_{0.18}$CoO$_{3}$ compound, shown (a) spin-up channel (top) and (b) spin-down channel (bottom).}
\label{}
  \vspace{-0.5cm}
\end{center}
\end{figure}
To examine the experimental data of $\alpha$ and $\sigma$, we have also calculated these transport coefficients corresponding to up and dn-spin channels. The values of $\alpha$ for up and dn-channels are shown in Fig.5a and 5b, respectively; whereas the values of $\sigma$ are shown in Fig. 5c and 5d, respectively. For the up-channel the values of $\alpha$ are small due to metallic nature, where as for dn-channel it is very large due to semiconducting nature. The calculated values of $\alpha$ for up-spin are $\sim$1.53 and $\sim$1.42 $\mu$V/K; whereas for dn-spin the values are $\sim$431 and $\sim$275 $\mu$V/K at 300 and 600 K, respectively. The values of $\sigma$/$\tau$ for up spin at 300 and 600 K are $\sim$16.13 $\times$10$^{19}$ $\Omega$$^{-1}$ m$^{-1}$ s$^{-1}$ and $\sim$16 $\times$10$^{19}$ $\Omega$$^{-1}$ m$^{-1}$ s$^{-1}$, respectively. In comparison to up-spin, the estimated values of $\sigma$ are very small for dn-spin channel and the calculated values of $\sigma$ for dn-spin at 300 and 600 K are found to be $\sim$0.13 $\times$10$^{17}$ $\Omega$$^{-1}$ m$^{-1}$ s$^{-1}$ and $\sim$4.4$\times$10$^{17}$ $\Omega$$^{-1}$ m$^{-1}$ s$^{-1}$, respectively. 

\begin{figure}[htbp]
  \begin{center}
   \includegraphics[width=0.40\textwidth]{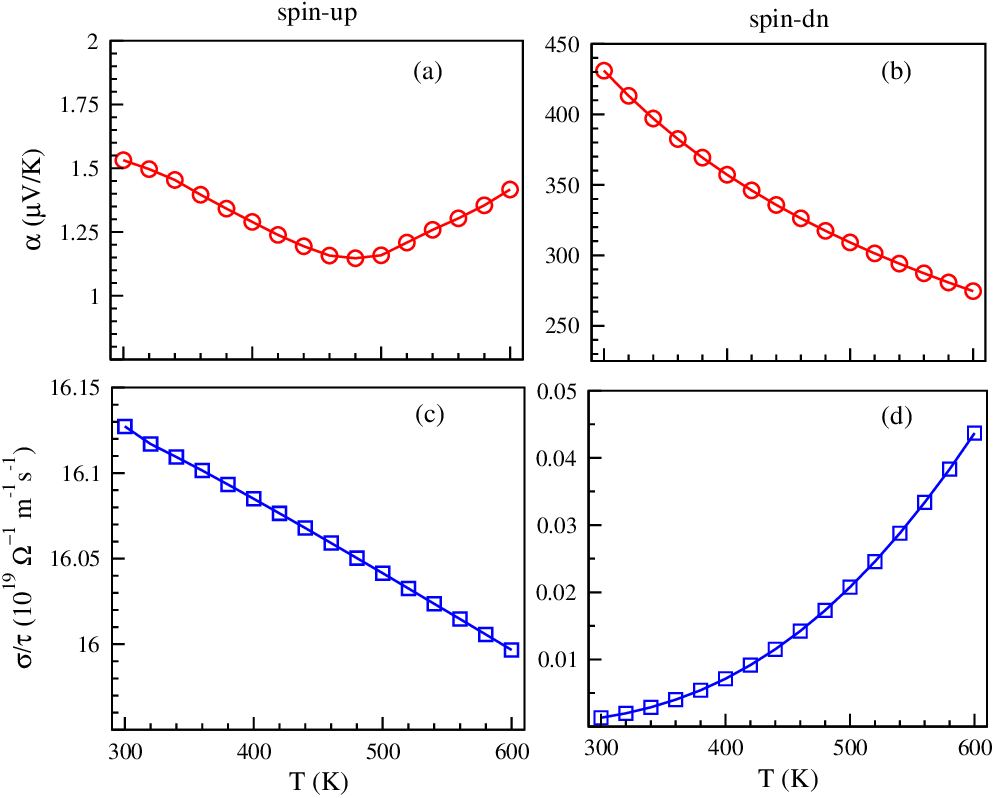}
    \label{}
    \captionsetup{justification=raggedright,
singlelinecheck=false
}
    \caption{(Color online) Variation of transport coefficients with temperature. (a and b) $\alpha$ with T, (c and d) $\sigma$ with T.}
    \vspace{-0.3cm}
  \end{center}
\end{figure}
In order to analyze the experimental values of the $\alpha$ and $\sigma$, the total values of these transport coefficients from up and dn-spin are need to be calculated. Therefore, two current model is used to calculate the total value of $\alpha$, where the values from up and dn-spin chanel is included. The expression of $\alpha$ given by two current model can be written as,\cite{Xiang, Botana}
\begin{equation}
    \alpha = [\frac{\alpha \uparrow (\sigma\uparrow/\tau_{up}(T)) + \alpha \downarrow (\sigma\downarrow/\tau_{dn}(T))}{\sigma\uparrow/\tau_{up}(T)+\sigma\downarrow/\tau_{dn}(T)}] 
  \end{equation}
  where, $\tau_{up}(T)$ and  $\tau_{dn}(T)$ are the used value of relaxation time for a given temperature T.
 The  Fig. 6a shows the temperature dependent plots of experimental and calculated values of $\alpha$ obtained by using the Eq$^{n}$ (1), where in the calculations of $\alpha$ we have considered equal and temperature independent $\tau$ ($\tau$$_{up}$ $\&$ $\tau$$_{dn}$). For the sake of clarity, calculated values of $\alpha$ vs T plot at zoomed scale is shown in the inset of the Fig. 6a, where we can clearly see that the values of $\alpha$ decreases very gradually upto $\sim$460 K and above that it increase sharply with further increase in temperature up to 600 K. The difference in the experimental and calculated values of $\alpha$ at 300, 480, and 600 K are $\sim$14.25, $\sim$10.12, and $\sim$12.42 $\mu$V/K, respectively. These differences are significantly large in the temperature range under study. Thus consideration of constant and equal values of $\tau$$_{up}$ and $\tau$$_{dn}$ are not suitable for providing a good match between experimental and calculated values of $\alpha$. In the 300-600 K range, scattering of free charge carriers through various interactions such as electron-electron and electron-phonon are expected to be more prominent and this results into change in the $\tau$. The change in values of $\tau$ plays an important role in the transport behavior of the system. Therefore, further analysis of experimental data have been carried out by considering the temperature dependent $\tau$ of up and dn-channel charge carriers.\\
\begin{figure}[htbp]
  \begin{center}
   \includegraphics[width=0.35\textwidth]{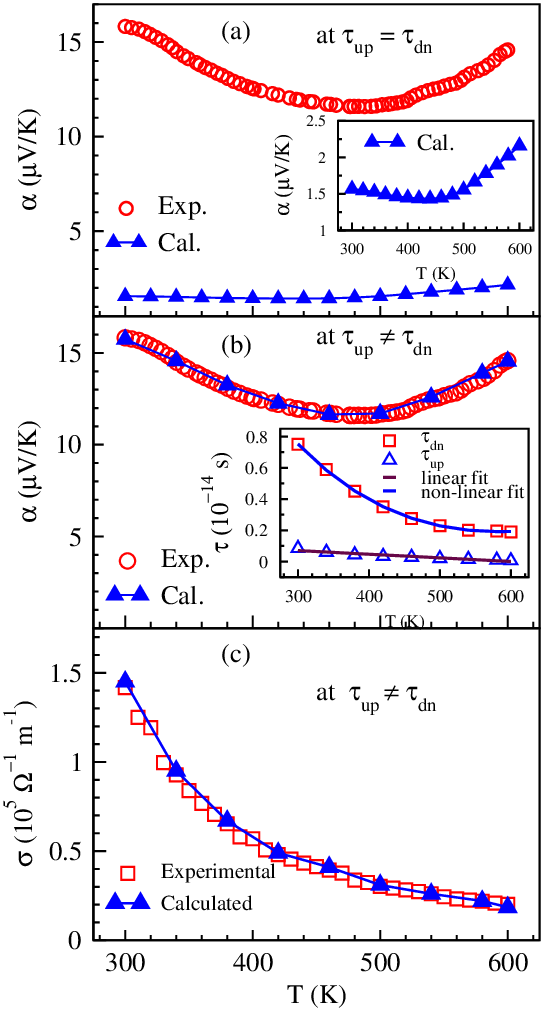}
    \label{}
    \captionsetup{justification=raggedright,
singlelinecheck=false
}
    \caption{(Color online) Variation of transport coefficients with temperature. Shown (a) and (b) $\alpha$ with temperature, (c) $\sigma$ with temperature.}
    \vspace{-0.3cm}
  \end{center}
\end{figure}
The temperature dependent values of $\tau$$_{up}$ and $\tau$$_{dn}$ have been chosen in such a way that a good match between experimental and calculated values of $\alpha$ were found in the 300-600 K range. Fig. 6b shows the combined plots of experimental and calculated values of $\alpha$, and the behavior of used values of $\tau$ for up and dn channel is also shown in the inset of this figure. The values of $\tau$$_{up}$ is almost linear, whereas non-linear variation in $\tau$$_{dn}$ is noticed. The used values of $\tau$$_{up}$ were $\sim$0.09 $\times$10$^{-14}$ s and $\sim$0.01 $\times$10$^{-14}$ s at 300 and 600 K, respectively. For dn-spin, values of $\tau$$_{dn}$ were taken $\sim$0.75 $\times$10$^{-14}$ s and $\sim$0.20 $\times$10$^{-14}$ s, respectively. Using these values of $\tau$ the calculated values of $\alpha$ at 300 and 600 K were found to be $\sim$15.75 $\mu$V/K and $\sim$14.55 $\mu$V/K, respectively. The calculated and experimental values are very closed to each other in the entire range of temperature under study. These adopted values of $\tau$ in present study for up and dn-spin are in the typical range, 10$^{-14}$-10$^{-15}$ s, reported for the metal and semiconductor.\cite{Ashcroft} In order to know the nature of variations in adopted values of $\tau$ with temperature, the best possible fitting were obtained by using linear equation (A$_{0}$ + A$_{1}$T) and cubic equation (B$_{0}$ + B$_{1}$T + B$_{2}$T$^{2}$ + B$_{3}$T$^{3}$) for $\tau$$_{up}$ and $\tau$$_{dn}$ values, respectively. The values of coefficients obtained from the fitting are A$_{0}$ ($\sim$0.14), A$_{1}$ ($\sim$ -2.3 $\times$10$^{-4}$), B$_{0}$ ($\sim$3.45), B$_{1}$ ($\sim$ -1.4 $\times$10$^{-2}$), B$_{2}$ ($\sim$ 1.96 $\times$ 10$^{-5}$) and B$_{3}$ ($\sim$ -8.68 $\times$ 10$^{-9}$). Further, the same set of $\tau$, which used in calculations of $\alpha$, were used to estimate the $\sigma$.\\
For the half-metallic system, $\sigma$ is normally sum of contributions from up and dn-spin channel and this can be defined as,\cite{Dorleijn}
\begin{equation}
  \sigma = \sigma\uparrow + \sigma\downarrow
\end{equation}
where, $\sigma\uparrow$ and $\sigma\downarrow$ are the electrical conductivities of the spin-up and spin-dn channels, respectively. With help of Eq$^{n}$(2) and using the values of $\sigma$ of up (Fig. 5c)and dn-spin (Fig. 5d), we have calculated the total $\sigma$ of the La$_{0.82}$Ba$_{0.18}$CoO$_{3}$. Fig. 6c shows the calculated and experimental values of $\sigma$. The calculated values of $\sigma$ are $\sim$1.45 $\times$ 10$^{5}$ $\Omega$$^{-1}$ m$^{-1}$ and $\sim$0.18 $\times$ 10$^{5}$ $\Omega$$^{-1}$ m$^{-1}$ at 300 and 600 K, respectively. These values are in very good agreement with the experimental data and shows that the adoption of $\tau$ values in calculation of $\alpha$ also capable independently in explaining the $\sigma$ data in 300-600 K range. At this point it is important to note that, in calculations of $\alpha$ and $\sigma$ we have used \textit{k-points} mesh and \textit{lpfac} parameter values equal to 125000
and 5, respectively. In the earlier studies it has been shown that transport coefficients above 300 K are not very much sensitive to the \textit{k-points} if the used value of \textit{k-points} in the Brillouin zone is equal to or greater than 64000.\cite{Sharma} Thus, we believe that the present calculations are capable in capturing of any small variations in the $\alpha$ and $\sigma$.\\
From the above discussions, the DFT+U calculations are appears to be successful in explaining the $\alpha$ and $\sigma$ data in the 300-600 K temperature range. For the compound considered in present study, there can be two possible reason behind the proper explanation of high temperature transport coefficients by DFT+U. (i) the effect of temperature dependent DOS around E$_{F}$ on transport coefficients are being adequately taken care through the temperature dependent $\tau$ used in calculations for $\alpha$ and $\sigma$. (ii) The temperature dependent DOS are not significant around E$_{F}$ (\textit{k$_{B}$}T range) for the temperature range under study, so that it do not affect much the values of transport coefficients. In a similar type of system, La$_{0.6}$Sr$_{0.4}$CoO$_{3}$, no significant temperature-dependent change in DOS is seen around E$_{F}$.\cite{Saitoh} The similar DOS feature can also be possible in the present compound. This can be possible reason for explaining the $\alpha$ and $\sigma$ data of La$_{0.82}$Ba$_{0.18}$CoO$_{3}$ in 300-600 K range using DFT+U method. If the first reason is behind the proper explanation of $\alpha$ and $\sigma$ data in present study then it will be interesting to see whether the same method can explain the TE data of other systems where change in DOS with temperature are significant around E$_{F}$.
\section{Conclusions} 
In conclusion, DFT+U calculations were found to be successful in understanding the TE properties of La$_{0.82}$Ba$_{0.18}$CoO$_{3}$ compound in 300-600 K temperature range. The non-monotic behavior of $\alpha$ were explained by considering the temperature dependent $\tau$ in two current model. The electrical conductivity of the system shows metallic behavior. The calculated values of $\sigma$ by using the same $\tau$, which explain the $\alpha$ data, also give the values close to experimentally observed data in 300-600 K range. The electronic structure calculations gives a FM ground state with half-metallic character having an energy gap of $\sim$270 meV for dn-spin channel. Applicability of DFT+U calculations have been found successful in explaining the high temperature TE behavior of La$_{0.82}$Ba$_{0.18}$CoO$_{3}$ system. It will also be interesting to see the versatility of this approach in understanding the high temperature TE behavior of other systems where temperature dependent changes in DOS around E$_{F}$ are significant. Therefore, in this directions a similar kind of studies on various other strongly correlated systems are desirable.

\end{document}